\documentclass[
aps,nofootinbib,
showpacs,showkeys,preprint
tightenlines,preprintnumbers,] {revtex4}

\usepackage{epsf,epsfig,subfigure,graphicx,amsmath,amssymb 
}
\usepackage{color}
\newcommand{\dis}[1]{\begin{equation}\begin{split}#1\end{split}\end{equation}}
\newcommand{\ie}{{\it i.e.~}}
\newcommand{\etal}{{\it et al.\,}}

\newcommand{\cred}[1]{{\color{red}{#1}}}

\newcommand{\darkZ}{dark-Z$'$}

\newcommand{\Mpt}{$M_{\rm P}$}

\newcommand{\Mg}{{M_{\rm GUT}}}
\newcommand{\Mgt}{$M_{\rm GUT}$} 
\newcommand{\Mi}{M_{\rm int}}
\newcommand{\Mit}{$M_{\rm int}$}
\newcommand{\vew}{$v_{\rm ew}$} 

\newcommand{\Uanom}{U(1)$_{\rm anom}$}

\newcommand{\SUflip}{SU(5)$\times$U(1)$_X$}

\newcommand{\Qanom}{Q_{\rm anom}}

\newcommand{\gev}{\,\textrm{GeV}}

\def\sw0{{$\sin^2\theta_W^0$}}

\newcommand{\Z}{{\bf Z}}

\def\smg{{SU(3)$_C\times
$SU(2)$_L\times$U(1)$_Y$}}
\def\tdz{{SU(2)$_{\rm dark}\times$U(1)$_Q$}} 
\def\E6{{\rm E_6}}

\def\EE8{{\rm E_8\times E_8'}}

\def\one{{\bf 1}}

\def\two{{\bf 2}}
\def\five{{\bf 5}}

\def\tenb{\overline{\bf 10}}

\def\fiveb{\overline{\bf 5}}

\begin{document}

\draft

\title{Naturally realized two dark  Z's near the electroweak scale } 
  
\author{Jihn E.  Kim}
\address
{Department of Physics, Kyung Hee University, 26 Gyungheedaero, Dongdaemun-Gu, Seoul 02447, Republic of Korea, and\\
Center for Axion and Precision Physics Research (IBS),
  291 Daehakro, Yuseong-Gu, Daejeon 34141, Republic of Korea
}
 
\begin{abstract} 
Chiral representations are the key in obtaining light fermions from some ultra-violet completed theories. The well-known  chiral example is one family set of fifteen chiral fields in the standard model.  We find a new chiral theory SU(2)$_{\rm dark}\times$U(1)$_Q$  with sixteen chiral fields, which does not have any gauge and gravitational anomalies.  
The group  SU(2)$_{\rm dark}\times$U(1)$_Q$ may belong to the dark sector, and we present a derivation of the spectrum from the E$_8\times$E$_8'$   hetrotic string. Necessarily, there appear two degrees at low energy: two   \darkZ's, or a \darkZ~plus a dark-photon. Being chiral, there is a chance  to probe this theory at TeV accelerators. Since the model belongs to the dark sector, the way to probe it is through the kinetic mixing.

\end{abstract}
\maketitle


\section{Introduction}\label{sec:Introduction}
 
In particle physics, there has been a deep question known as the gauge hierarchy problem: ``How do  the SM fermions appear at such a small  electroweak scale, compared to an ultra-violet completed scale,  the Planck mass \Mpt~or the grand unification (GUT) scale \Mgt ?''   Two issues in the  hierarchy problem are: (i) obtaining massless SM particles at the  ultra-violet completed scale, and (ii) rendering the elctroweak scale masses to the SM fermions. The first issue is resolved by the profound and simple requirement,  a chiral theory at the ultra-violet completed scale \cite{Georgi79}. The second issue is the method obtaining the vacuum expectation value (VEV) of the Higgs field at the electroweak scale \vew$\simeq 246\,\gev$, a kind of TeV scale, for which the most well-known example is supersymmetry (SUSY) \cite{Ibanez82}.
 
In this paper, we propose that any particles appearing at the TeV scale, for a detection possibility at the  LHC, must satisfy Condition (i).  The best known example is a spinor representation in the SO($4n+2$) GUT models \cite{SpinorSO}. Orbifolding in extra dimensions presents a possibility of massless particles, as shown in a simple field theoretic orbifold \cite{Kawa01}. But, the orbifold compactification in string theory is the prototype example \cite{OrbString}, providing a simple geometrical interpretation. 
Note, however, that fermionic constructions \cite{fermionic} and Gepner models \cite{Gepner} have been also used in four dimensional (4D) phenomenology from  string. In these 4D constructions, it was necessary to check whether vectorlike representations of exotically charged particles, which appear quite often, are present or not as studied in Refs. \cite{Exotics}.
 
Anyway, Condition (i) is the basic requirement we satisfy at low energy effective theory in 4 dimensions (4D). To realize Condition (ii),  model parameters are required to be known in detail, and hence we do not discuss it here except pointing out several mass scales in particle physics.
The SM a chiral theory, realized in Nature,  describes  the electroweak scale physics  successfully. So, we anticipate that if a natural chiral model is found then it might have a great chance to be realized in Nature. Since any new particle has not been detected at the LHC so far, a new particle in the new chiral theory,  which interacts with the SM sector extremely feebly, must be in the dark sector. Here, the dark sector is not introduced just for explaining cold dark matter (CDM) of the Universe. The well-known CDM examples, ``invisible'' axions \cite{KSVZ} and weaky interacting massive particles (WIMPs) \cite{LeeWein77}, are belonging to the visible sector. On the other hand, the heterotic $\EE8$ string \cite{Gross84} implies a possibility of dark sector from E$_8'$. If the dark sector introduces CDM, then it is just a bonus.\footnote{In this paper, we do not introduce the dark sector for the sake of dark matter \cite{Arkani09}.}
Even though the dark sector interacts with the SM sector extremely feebly, it can be probed by the kinetic mixing terms \cite{Okun:1982xi} of two SM gauge bosons, photon and $Z$. Since the dark sector does not carry the SM weak hyperchage, the charge raising and lowering gauge bosons in the dark sector cannot have kinetic mixings with $W^\pm$ of the SM. 

We will show that the simplest chiral theory which does not have any gravitational and gauge anomalies is \tdz. Because the rank of this gauge group is 2, we will have two $Z'$ gauge bosons at low energy. This minimal model will be called Two Dark $Z$  model (TDZ).  The first new particles observed at the  CERN SPS proton-antiproton collider were $W^\pm$ and $Z$
\cite{WZdiscover}. This is because it is relatively easy to identify leptons at high energy colliders. With this new chiral model, therefore, we expect that the first new particles expected at the LHC are two dark $Z$'s.

In Sec. \ref{sec:Minimal}, we present the minimal chiral model. In Sec. \ref{sec:KinetcMix}, the kinetic mixing in the \tdz~is discussed.  In Sec. \ref{sec:String}, it is shown that the minimal chiral model is derivable from a string compactification. Section \ref{sec:Conclusion} is a conclusion. In Appendix, a SUSY scenario based on the hidden sector SU(5)$'$ from $\Z_{12-I}$ orbifold compactification \cite{Huh09} is discussed.

\section{Minimal chiral model}\label{sec:Minimal}

If a new gauge boson pops up near the electroweak scale, it must be from a chiral theory. A chiral theory near the electroweak scale should not have gravititonal and gauge anomalies. First, consider the rank-1 gauge groups. If we consider an SU(2), it is not possible to have a chiral theory because there must be even number of doublets \cite{WittenSU2}. With only one U(1)$_Y$ group, the absence of gravitational anomaly requires Tr\,$Y=0$ and  the absence of gauge anomaly in addition requires Tr\,$Y^3=0$. For example, even though two charged fields $Y=+1$ and $-1$ do not have these anomalies, the model is not allowed in our framework because it is vector-like.   But if we use the $Y$ of the SM, these two conditions are satisfied.\footnote{In addition, the SM  requires Tr\,$Y=0$ for quarks and leptons separately, and also additional conditions for the absence of non-Abelian gauge anomalies.} 
Second, let us consider the rank-2 gauge groups,
\dis{
&{\rm SU(3):~Vectorlike~ representations,~ hence~not~allowed,}  \\
&{\rm SU(2) \times SU(2):~No~ chiral~ theory~with~even~number~of~doublets,}  \nonumber \\
&{\rm U(1) \times U(1)':~Six~ conditions~ for~the~absence~of~anomalies},~\{{\rm Tr}Y,{\rm Tr}Y',{\rm Tr}Y^3,{\rm Tr}Y^{\prime 3},{\rm Tr}YY^{\prime 2},{\rm Tr}Y^2Y'\}=0, 
 }
 and
\begin{eqnarray}
  {\rm SU(2) \times U(1) :~Two~ conditions~ with~doublets~and~singlets },~\{{\rm Tr}Y, {\rm Tr}Y^3\}=0.
\label{eq:Conditions}
\end{eqnarray}
 Thus, the  simplest case  is \tdz, and at least two \darkZ's are predicted at low energy.  Two conditions in (\ref{eq:Conditions}) for $N$(=\,even) doublets and $2N$ singlets are 
\dis{
&\sum_{i=1}^{4N}\,Q_i=0 , \label{eq:Tr1}
}
 and
 \dis{
&\sum_{i=1}^{4N}\,Q_i^3=\left(\sum_{i=1}^{4N}\,Q_i \right)\left( \sum_{i=1}^{4N}\,Q_i^2-\sum_{i\ne j}^{4N}\,Q_iQ_j \right)+   3\sum_{i\ne j\ne k}^{4N}\,Q_iQ_jQ_k =0 .\label{eq:Tr3}
}
Condition (\ref{eq:Tr1}) satisfies Eq. (\ref{eq:Tr3}) if the term $\sum_{i\ne j\ne k}$  is vanishing. The number of terms in this sum is 
\dis{
 \begin{pmatrix}4N\\ 3 \end{pmatrix} 
}
which is very large. So, a complete search is more involved. The well-known chiral theory, satisfying  (\ref{eq:Tr3}), is the SM, \smg. 

Here, we present a simpler one \tdz, satisfying the conditions in  (\ref{eq:Tr1},\ref{eq:Tr3}), with the following fermions
\dis{ 
&Q=\frac12:\quad \ell_i=\begin{pmatrix} E_{i}\\[0.6em] N_{i}
\end{pmatrix}_{\frac{+1}{2}},~\begin{array}{c}E_{i\,,-1}^c\\[0.6em]  N_{i\,,0}^c\end{array},
~(i=1,2,3)\\[0.7em]
&Q=-\frac{3}2:~ {\cal L}=\begin{pmatrix} {\cal E} \\[0.6em] {\cal F} 
\end{pmatrix}_{\frac{-3}{2}},~~\begin{array}{c}{\cal E}_{,+1}^c\\[0.6em]  
{\cal F}_{ ,+2}^c\end{array} 
\label{eq:MinModel}
 }
where the subscripts denote the $Q$ charges. There are four doublets without the SU(2) anomaly \cite{WittenSU2}. One set, one of $E_{i\,,-1}^c$ and ${\cal E}_{,+1}^c$, forms a vector-like pair, but we keep them to provide masses for all the particles after breaking \tdz. In Eq. (\ref{eq:MinModel}), there appear 16 left-handed chiral fields,\footnote{Note that the SM has 45 chiral fields.} and they do not introduce any gravitational and gauge anomalies. It is interesting to observe that there appear 16 chiral fields as in the spinor representation {\bf 16} in SO(10).  Eq.  (\ref{eq:MinModel}) realizes the TDZ model.

To break the rank-2 gauge group \tdz~completely and to give fermions masses, let us introduce  two doublets\footnote{With supersymmetry, we need two doublets to make all chiral fields massive.} and a singlet of scalars,
 \dis{ 
 \Phi_u=\begin{pmatrix} \phi_u^+\\[0.6em] \phi_u^0
\end{pmatrix}_{Q=\frac{+1}{2}},~ \Phi_d=\begin{pmatrix} \phi_d^0\\[0.6em] \phi_d^-
\end{pmatrix}_{Q=\frac{-1}{2}},~S_{Q=2} ,
\label{eq:Higgs}
 }
where their  vacuum expectation values (VEVs) are,\footnote{Choosing the $Q=2$ singlet for breaking U(1) is just for an illustration.
}
\begin{eqnarray}
\langle \phi_{u,d}^0\rangle   =\frac{V_{u,d}}{\sqrt2},~\langle S\rangle   =\frac{V_{S}}{\sqrt2}.\label{eq:DSvevs}
\end{eqnarray}
Then,  masses of two \darkZ's are
\begin{eqnarray} 
&&M^2_{Z'_1}= ({g_2^2+g_Q^2}) V_D^2=\frac{g_2^2}{\cos^2\theta}\,V_D^2,\label{eq:gaugeZ} 
\\[0.5em] 
 &&M^2_{Z'_2}=(2g_Q)^2\frac{V_S^2}{2}=2g_Q^2V_S^2=2g_2^2\tan^2\theta\, V_S^2,\label{eq:gaugeA} 
\end{eqnarray} 
where $g_2$ and $g_Q$ are the SU(2)$_{\rm dark}$ and   U(1)$_Q$ couplings, respectively,  $\tan\theta\equiv g_Q/g_2$,  and $V_D^2=V_{u}^2+V_{d}^2$. Thus, the mass ratio of two \darkZ~masses is
\begin{eqnarray} 
  r=  \sqrt{2}\left|\frac{V_S}{V_D}\,\sin\theta \right|.
\label{eq:ratio} 
\end{eqnarray}  
If $V_S\to 0$, $Z'_2$ may be called dark-photon, which is included in our terminology  TDZ. This estimate will be used in Appendix. 

\section{The kinetic mixing}\label{sec:KinetcMix}

If multiple dark-U(1) gauge bosons are present, they can mix with the SM photon, 
 most probably via kinetic mixings as suggested in \cite{Okun:1982xi}. Since the rank of \tdz\,gauge group is 2, there are two \darkZ's and we summarize their kinetic mixing  with photon.\footnote{Their  mixing with the $Z$-boson is omitted here.} These arise   via loops between photon and dark-photon through an intermediate particle(s) $\chi$ which carries both the electromagnetic and dark charges.   After a proper
diagonalization procedure of the kinetic energy terms, then the
electromagnetic charge of $\chi$ can be millicharged, $ {\cal O}(\alpha/2\pi)e$. In the heterotic $\EE8$
string model, the extra  E$_8'$ gauge group may contain dark-photons which will be called \darkZ's, leading to the kinetic mixing of  $ {\cal O}(\alpha/2\pi)$\cite{Dienes97}. Indeed, an
explicit model for this kind from string compactification exists in the literature~\cite{KimJE:2007,Cheung07}.

The intermediate ${\cal O}$(MeV) millicharged particles have not been ruled out by observations in the previous study~\cite{Davidson:2000hf}.   For the discovery possibility at the LHC, we consider the electroweak scale   \darkZ's.

Consider three Abelian gauge groups U(1)$_{\rm QED}$ and  U(1)$_{i}\,(i=1,2)$.  The kinetic mixing of  U(1)$_{\rm
QED}$  and  U(1)$_{i}$ \darkZ,s are parameterized by, following the notation of \cite{ParkJC08},
\begin{eqnarray}
{\cal L} = -\frac{1}{4}\hat{F}_{\mu\nu}\hat{F}^{\mu\nu}
-\frac{1}{4} \hat{X}^1_{\mu\nu}\hat{X}^{1\,\mu\nu}-\frac{1}{4} \hat{X}^2_{\mu\nu}\hat{X}^{2\,\mu\nu}
- \frac{\xi_1}{2}\hat{F}_{\mu\nu}\hat{X}^{1\,\mu\nu} 
- \frac{\xi_2}{2}\hat{F}_{\mu\nu}\hat{X}^{2\,\mu\nu} 
- \frac{\xi_{12}}{2}\hat{X}^1_{\mu\nu}\hat{X}^{2\,\mu\nu},  \label{lagrangian}
\end{eqnarray}
where  $\hat{A}_\mu (\hat{X^i}_{\mu})$ is the  U(1)$_{\rm QED}$
(dark--U(1)$^{i}$) gauge boson, and its field strength tensor is
$\hat{F}_{\mu\nu} (\hat{X}^i_{\mu\nu})$. The kinetic mixings are
parameterized by $\xi$'s which are generically allowed by the gauge invariance and the Lorentz symmetry. In the low-energy effective theory, $\xi$'s are considered to be completely arbitrary
parameters. An ultraviolet-completed theory is expected to generate the kinetic mixing parameters. The usual
diagonalization procedure of these kinetic terms leads to the
relation,
\begin{eqnarray}
\left(
  \begin{array}{c}
    A_\mu \\[0.5em]
    X^1_\mu \\[0.5em]
   X^2_\mu \\
  \end{array}
\right) = \left(
 \begin{array}{ccc}
   B_{11}    & 0 &~0\\  [0.5em] 
  \frac{\xi_1-\xi_2\xi_{12}}{\sqrt{1-\xi_{12}^2}} &  \sqrt{1-\xi_{12}^2} &~0\\ [0.5em]
  \xi_2 &\xi_{12}&~1
 \end{array}
  \right) \left(
  \begin{array}{c}
   \hat{A}_\mu \\[0.5em]
 \hat{X}^1_\mu \\[0.5em]
 \hat{X}^2_\mu\end{array}
 \right),\label{eq:transf1}
\end{eqnarray}
where
\begin{eqnarray}
B_{11}=
   \sqrt{1-\frac{(\xi_1-\xi_2)^2+2\xi_1\xi_2(1-\xi_{12})}{ 1-\xi_{12}^2 }} 
\end{eqnarray}
and we obtain
\begin{align}
{\cal L} =
-\frac{1}{4}F_{\mu\nu}F^{\mu\nu}-\frac{1}{4}X^1_{\mu\nu}X^{1\,\mu\nu}-\frac{1}{4}X^2_{\mu\nu}X^{2\,\mu\nu},
\end{align}
where the new field strengths are  $F_{\mu\nu}, X^1_{\mu\nu}$, and $X^2_{\mu\nu}$.
Photon corresponds to $A_\mu$ and { \darkZ s} correspond to
$X^i_\mu\,(i=1,2)$. If the \darkZ's are exactly massless, there exists an
 SO(3) symmetry in the $A_\mu-X^i_\mu$ field space. 
 
Using the above SO(3) symmetry, let us take the following simple
interaction Lagrangian of a SM fermion with a
photon in the original basis as
\begin{align}
{\cal L}= \bar{\psi} \left( Q\hat{e} \, \gamma^\mu \right) \psi
\hat{A}_\mu. \label{SM-interaction}
\end{align}
Note that in this basis there is no direct interaction between the
electron and the hidden sector gauge boson $\hat{X}$. If there
exists a hidden sector Dirac fermion $\chi$ with the 
U(1)$_{\rm ex}$ charge $Q_{\chi}$, its interaction with the hidden sector
gauge boson is simply represented by
\begin{align}
{\cal L}= \bar{\chi} \left( \hat{e}^{\rm ex}_i Q^{\rm ex}_i  \gamma^\mu
\right) \chi \hat{X}^i_{\mu}, \label{hidden-interaction}
\end{align}
where $\hat{e}_{\rm ex}$ can be different from $\hat{e}$ in
general. In this case, there is also no direct interaction between
the hidden fermion and the visible sector gauge boson $\hat{A}_\mu$.
We can recast the Lagrangian~(\ref{SM-interaction}) in the
transformed basis $A_\mu$ and $X_\mu$,
\begin{eqnarray}
{\cal L}= \bar{\psi} \left( \frac{\sqrt{1-\xi_{12}^2}}{\sqrt{ 1-\xi_1^2-\xi_{12}^2+2\xi_1\xi_2\xi_{12}}} \,Q\hat{e}\right) \gamma^\mu  \psi A_\mu,
\label{eq:photonCoupl}
\end{eqnarray}
where we used the inverse of (\ref{eq:transf1})
\begin{eqnarray}
\frac{1}{\rm Det}
\begin{pmatrix}
\sqrt{1-\xi_{12}^2},&0,&0 \\
- \frac{\xi_1-\xi_2\xi_{12}}{\sqrt{1-\xi_{12}^2}},&B_{11},&0\\
- \frac{\xi_2-\xi_1\xi_{12}}{\sqrt{1-\xi_{12}^2}},&-\xi_{12}
B_{11},&B_{11}\sqrt{1-\xi_{12}^2}
\end{pmatrix}\label{eq:transf2}
\end{eqnarray}
with 
\begin{align}
{\rm Det}=\sqrt{ 1-\xi_1^2-\xi_2^2-\xi_{12}^2+2\xi_1\xi_2\xi_{12}}.\nonumber
\end{align}
Here, one notices that the standard model fermion has a coupling
only to the visible sector gauge boson $A_\mu$ even after changing the basis of the gauge bosons. However, the coupling constant
$\hat{e}$ is modified to $e$ as suggested in Eq. (\ref{eq:photonCoupl}). 
Similarly, we derive the following couplings for
$\chi$,
\begin{align}
{\cal L} = \bar{\chi} \gamma^\mu  
 \left[\frac{\hat{e}^{\rm ex}_1}{\rm Det}  \left(B_{11}X^1_\mu
- \frac{\xi_1-\xi_2\xi_{12}}{\sqrt{1-\xi_{12}^2}}   A_\mu\right) Q_1^{\rm ex}
+ \frac{\hat{e}^{\rm ex}_2}{\rm Det}  \left(B_{11}\sqrt{1-\xi_{12}^2} X^2_\mu
- {B_{11}\xi_{12}}  X^1_\mu -\frac{\xi_2-\xi_1\xi_{12}}{\sqrt{1-\xi_{12}^2}} A_\mu\right)Q_2^{\rm ex} \right] \chi .
\label{hidden-shift}
\end{align}
In this basis, the hidden sector matter field $\chi$ now can
couple to the visible sector gauge boson $A_\mu$ with the couplings
$-\hat{e}^{\rm ex}_1Q_1^{\rm ex}( \xi_1 - \xi_2\xi_{12})/\sqrt{1-\xi_{12}^2} /{\rm Det}$ to the mass eigenstate $X^1_\mu$ and $\hat{e}^{\rm ex}_2Q_2^{\rm ex}( \xi_2 - \xi_1\xi_{12})/\sqrt{1-\xi_{12}^2} /{\rm Det}$ to the mass eigenstate $X^2_\mu$. In terms of the
afore-mentioned  SO(3) symmetry, it simply means the mismatch
between the gauge couplings of the electron and other fermions.
Thus, we can set the physical hidden sector coupling $e_{\rm ex}$
as $e_{\rm ex} \equiv \hat{e}_{\rm ex}$ and we define the coupling
of the field $\chi$ to the visible sector gauge boson $A_\mu$,
introducing the millicharge parameter $\varepsilon_i$, as
$\varepsilon_i e $ such that
\begin{eqnarray}
 e=\frac{\hat{e}}{\sqrt{1-\xi_1^2-\xi_2^2}},&& \varepsilon_1=-\frac{\hat{e}^{\rm ex}_1}{e}\frac{\xi_1-\xi_2\xi_{12}}{\sqrt{  1-\xi_1^2-\xi_2^2-\xi_{12}^2+2\xi_1\xi_2\xi_{12}}}\approx -\frac{\hat{e}^{\rm ex}_1}{e} \xi_1,\\ && \varepsilon_2=-\frac{\hat{e}^{\rm ex}_2}{e}\frac{\xi_2-\xi_1 \xi_{12} }{\sqrt{  1-\xi_1^2-\xi_2^2-\xi_{12}^2+2\xi_1\xi_2\xi_{12}}}\approx -\frac{\hat{e}^{\rm ex}_2}{e} \xi_2.
\end{eqnarray}
Note in general that $e\ne e^{\rm ex}_i$. Since $\xi_{i,12} \simeq {\cal O}(\varepsilon_i
e/e^{\rm ex}_i)$ is expected to be small, the condition $\xi_{i,12}<1$ gives $\alpha^{\rm ex}_i/{\alpha} > \varepsilon^2$. From a fundamental
theory, one can calculate the ratio $e_{\rm ex}/e$ in principle,
which is possible with the detail knowledge of the
compactification radius~\cite{KKmasses}.  

Similarly, one can calculate the mixing of the SM $Z$ boson with \darkZ's, which however is not presented here.

\section{From a string model} \label{sec:String}

In this section, we derive the minimal chiral model (\ref{eq:MinModel}) discussed in Sec. \ref{sec:Minimal} from a string theory. The $\EE8$ heterotic string model compactified on $\Z_{12-I}$ orbifold gives  the flipped SU(5)$_{\rm flip}$ times SU(5)$'\times$SU(2)$'$ with several extra U(1)'s \cite{Huh09}. Here, the factor  SU(5)$_{\rm flip}$ contains a gauge group U(1): \SUflip. The first important U(1) gauge group is U(1)$_X$ in \SUflip, which is free of any gauge anomaly. The second is the anomalous  \Uanom.  Except these two U(1) factors, U(1)$_X$ and \Uanom, the non-Abelian gauge group is SU(5)$\times$SU(5)$'\times$SU(2)$'$.  Note that the charges of  U(1)$_X$ and \Uanom~are,\footnote{For the definition, see, Ref. \cite{KimKyae17}.}
\begin{eqnarray}
X=(-2,-2,-2,-2,-2\,;\,0^3)(\,0^8)'.\label{eq:X} 
\end{eqnarray}
\begin{eqnarray}
  \Qanom= 84Q_1+147Q_2 -42Q_3-63Q_5- 9Q_6,\label{eq:Qanom}
\end{eqnarray}
where
\begin{eqnarray}
 Q_1&=&(0^5;12,0,0)(0^8)',\nonumber \\[0.3em]
 Q_2&=&(0^5;0,12,0)(0^8)',\nonumber\\[0.3em]
 Q_3&=&(0^5;0,0,12)(0^8)',\nonumber\\[0.3em]
 Q_4&=&(0^8)(0^4,0;12,-12,0)',\nonumber\\[0.3em]
 Q_5&=&(0^8)(0^4,0;-6,-6,12)',\nonumber\\[0.3em]
Q_6&=& (0^8)(-6,-6,-6,-6,18;0,0,6)'. \nonumber
\end{eqnarray}

In the orbifold compactification, frequently there appears an anomalous U(1)$_A$ gauge fields $\tilde{A}_\mu$ from a subgroup of $\EE8$ \cite{Anom87}. The charge of this anomalous  U(1)$_A$ is given in Eq. (\ref{eq:Qanom}).  In addition, the anomaly cancellation in ten dimensions (10D) requires the so-called Green-Schwarz (GS) term in terms of the second rank antisymmetric-tensor field $B_{MN}\,(M,N=1,2,\cdots,10)$ \cite{GS84}. The 10D $B_{MN}$ always introcuces a model-independent (MI) axion  $a_{\rm MI}$ in 4D, $\partial_\mu a_{\rm MI}\propto \epsilon_{\mu\nu\rho\sigma}
H^{\nu\rho\sigma}\,(\mu,\rm etc.=1,2,3,4)$ where $H^{\nu\rho\sigma}$ is the field strength of $B^{\rho\sigma}$  \cite{Witten84}. The anomalous U(1) gauge boson absorbs the MI axion to become massive and there results a global symmetry \Uanom~below the compactification scale.  More phenomenologically, \Uanom~can be suggested for a plausible flavor symmetry \cite{KimKyae17}.  The global symmetry \Uanom~is good for a Peccei-Quinn symmetry \cite{PQ77} toward ``invisible'' axions at the intermediate scale $\Mi$ \cite{KSVZ}.  Except the two U(1)'s, Eqs. (\ref{eq:X},\ref{eq:Qanom}), all U(1)'s are assumed to be  broken at a high energy scale, much above \Mit.  
In a more detail, it works as follows.  Suppose that five U(1) charges out of $Q_{1,\cdots,6}$ are broken, and there is only one gauge symmetry remaining, which we identify as \Uanom.   Now, we can consider two continuous parameters, one is the MI-axion direction and the other the phase of \Uanom~transformation. Out of two continuous directions, only one phase or pseudoscalar is absorbed by the \Uanom~gauge boson, and one continuous direction survives.
The remaining continuous degree corresponds to a global symmetry, which  is called the 't Hooft mechanism \cite{Hooft71}: ``If both a gauge symmetry and a global symmetry are broken by one scalar VEV, the gauge symmetry is broken and a global symmetry is surviving''.  The resulting global charge is a linear combination of the original gauge and global charges.   Even though we obtain a global symmetry \Uanom, it is obtained from the original two gauge symmetries, one from the two-index anti-symmetric tensor gauge field $B_{MN}$ in 10D and the other the \Uanom~subgroup of $\EE8$ given in Eq. (\ref{eq:Qanom}).
 
 Here, the primed goups  SU(5)$'\times$SU(2)$'$ are the hidden sector non-Abelian gauge groups. The hidden sector representations under SU(5)$'\times$SU(2)$'$ are given in Tables \ref{tb:SUhfields} and \ref{tb:SU2fields} \cite{KimKyae17}. 
 After removing vector-like representations from Tables \ref{tb:SUhfields} and \ref{tb:SU2fields}, we obtain
\dis{
 &(\overline{\bf 10}',{\bf 1}'),~( {\bf 5}',{\bf 2}'),~(\overline{\bf 5}',{\bf 1}'),~({\bf 1}',{\bf 2}'),~~ {\rm under~ SU(5)'}\times{\rm  SU(2)'}.\\
 &\to  \Psi_{AB} ,~  \Phi^{A\alpha},~   \psi_{A} ,~
 \phi^{\alpha} , \label{eq:HidReps}
 }
where the tensor notation is used in the second line with the SU(5)$'$ index $A=\{1,2,3,4,5\}$ and the SU(2)$'$ index $\alpha=\{1,2\}$. The representations in (\ref{eq:HidReps}) do not lead to an SU(5)$'$ anomaly.
  Let two SU(2) subgroup indices of  SU(5)$'$ be $i=\{1,2\}$ and $I=\{4,5\}$ so that the five SU(5)$'$ indices  split into 
    \begin{eqnarray}
\{A\}\equiv  \{ i,3,I\}.
\end{eqnarray}
 
\begin{table}[!t]
{\tiny
\begin{center}
\begin{tabular}{|c|c|c|c||ccccccc|c|c}
\hline Sect. & States & SU(5)$'$ & Multiplicity & $Q_1$& $Q_2$ & $Q_3$ & $Q_4$ & $Q_5$ & $Q_6$ &  $Q_{\rm anom}$ & Label    \\[0.3em]
\hline\hline

$T_{1}^0$ &
$\left(0\,0\,0\,0\,0\,;\frac{-1}{6}\,\frac{-1}{6}\,\frac{-1}{6}\,\right)
(\underline{-1\,0\,0\,0\,}\,0\,; \frac{1}{4}\,\frac{1}{4}\,\frac{1}{2} )'$ &  $\tenb'_0$  & $1$ &$-2$ &$-2$ &$-2$ &0 &$+3$ & $+9$ &$-648$\cred{(-$\frac{36}{7}$)}  & $ T_1'$    \\[0.5em]
 & $\left(0\,0\,0\,0\,0\,;\frac{-1}{6}\,\frac{-1}{6}\,\frac{-1}{6}\,\right)
(\underline{\frac{1}{2}\,\frac{1}{2}\,\frac{-1}{2}\,\frac{-1}{2}\,}\,\frac{1}{2}\,; \frac{-1}{4}\,\frac{-1}{4}\,0 )'$ &      &&&&&&&& &  \\[0.5em]
\hline
$T_{1}^0$ &$\left(0\,0\,0\,0\,0\,;\frac{-1}{6}\,\frac{-1}{6}\,\frac{-1}{6}\,\right)
(\underline{-1\,0\,0\,0\,}\,0\,; \frac{1}{4}\,\frac{1}{4}\,\frac{1}{2} )'$ &  $(\five',\two')_0$   & $1$ &$-2$ &--2 &--2 &0  &$+3$ &$+9$ &$-648$\cred{(-$\frac{36}{7}$)}  & $ F_1'$ \\[0.2em]
 &$\left(0\,0\,0\,0\,0\,;\frac{-1}{6}\,\frac{-1}{6}\,\frac{-1}{6}\,\right)
({\frac12\,\frac{1}{2}\,\frac{1}{2}\,\frac{1}{2}\,}\,\frac{-1}{2}\,\,; \frac{-1}{4}\,\frac{-1}{4}\,0 )'$  && &&&&&&&& \\[0.2em]
&$\left(0\,0\,0\,0\,0\,;\frac{-1}{6}\,\frac{-1}{6}\,\frac{-1}{6}\,\right)
(\underline{\frac12\,\frac{-1}{2}\,\frac{-1}{2}\,\frac{-1}{2}\,}\,\frac{-1}{2}\,\,; \frac{-1}{4}\,\frac{-1}{4}\,0 )'$ &  &&&&&&&& &    \\[0.2em]
&$\left(0\,0\,0\,0\,0\,;\frac{-1}{6}\,\frac{-1}{6}\,\frac{-1}{6}\,\right)
({0\,0\,0\,0\,}\,0\,; \frac{-3}{4}\,\frac{-3}{4}\,\frac{-1}{2} )'$ &  &&&&&&&& &  \\[0.6em]
\hline
$T_{1}^0$&$\left(0\,0\,0\,0\,0\,;\frac{-1}{6}\,\frac{-1}{6}\,\frac{-1}{6}\,\right)
(\underline{\frac{-1}{2}\,\frac{1}{2}\,\frac{1}{2}\,\frac{1}{2}\,}\,\frac{-1}{2}\,\,; \frac{-1}{4}\,\frac{-1}{4}\,0 )'$ & $\fiveb'_0$  & $1$ &--2 &--2 &--2 &0 &$+3$ &$-15$ &$-432$\cred{(-$\frac{24}{7}$)}  &  $F_2'$   \\[0.2em]
&$\left(0\,0\,0\,0\,0\,;\frac{-1}{6}\,\frac{-1}{6}\,\frac{-1}{6}\,\right)
(0\,0\,0\,0\,-1 \,; \frac{1}{4}\,\frac{1}{4}\,\frac{1}{2} )'$ && &&&&&&&&  \\[0.3em]
\hline

$T_{1}^+$
&$\left(\frac{1}{6}\,\frac{1}{6}\,\frac{1}{6}\,\frac{1}{6}\,\frac{1}{6}\,;\frac{1}{3}\,
\frac{-1}{3}\,0 \,\right)
\left(\underline{\frac{-5}{6}\,\frac{1}{6}\,\frac{1}{6}\,\frac{1}{6}\,}\,\frac{1}{2}\,
;\frac{1}{12}\,\frac{-1}{4}\,0\,\right)'$ &  $\fiveb'_{-5/3}$   & $1$ &$+4$ &$-4$ &0 &$+4$
&$+1$ &$+11$ &$-414$\cred{(-$\frac{23}{7}$)}  & $F'_3$  \\[0.2em]
&$\left(\frac{1}{6}\,\frac{1}{6}\,\frac{1}{6}\,\frac{1}{6}\,\frac{1}{6}\,;\frac{1}{3}\,
\frac{-1}{3}\,0 \,\right)
\left(\frac{-1}{3}\,\frac{-1}{3}\,\frac{-1}{3}\,\frac{-1}{3}\,0\,;\frac{7}{12}\,
\frac{1}{4}\,\frac{1}{2}\,\right)'$ &&&&&&&&&&     \\[0.3em]
\hline

$T_{4}^+$ &
$\left(\frac{1}{6}\,\frac{1}{6}\,\frac{1}{6}\,\frac{1}{6}\,\frac{1}{6}\,;\frac{-1}{6}\,
\frac{1}{6}\,\frac{1}{2}\,\right)
\left(\underline{\frac{2}{3}\,\frac{-1}{3}\,\frac{-1}{3}\,\frac{-1}{3}\,}\,0\,;
\frac{1}{3}\,0\,0\,\right)'$ &  $\five'_{-5/3}$  & $3$ &--2 &+2 &+6 &$+4$ &$-2$ &$+2$ &$-18$\cred{(-$\frac{1}{7}$)}  & $ F'_4$  \\[0.2em]
 &$\left(\frac{1}{6}\,\frac{1}{6}\,\frac{1}{6}\,\frac{1}{6}\,\frac{1}{6}\,;
 \frac{-1}{6}\,\frac{1}{6}\,\frac{1}{2}\,\right)
\left(\frac{1}{6}\,\frac{1}{6}\,\frac{1}{6}\,\frac{1}{6}\,\frac{1}{2}\,;
\frac{-1}{6}\,\frac{-1}{2}\,\frac{-1}{2}\,\right)'$ & && &&&&&& &     \\[0.2em]
\hline
$T_{4}^-$ &
$\left(\frac{-1}{6}\,\frac{-1}{6}\,\frac{-1}{6}\,\frac{-1}{6}\,\frac{-1}{6}\,;
\frac{-1}{6}\,\frac{-1}{2}\,\frac{1}{6}\,\right)
\left(\underline{\frac{-2}{3}\,\frac{1}{3}\,\frac{1}{3}\,\frac{1}{3}\,}\,0\,;
\frac{-1}{3}\,0\,0\,\right)'$ &  $\fiveb'_{5/3}$   & $3$ &--2 &--6 &+2 &$-4$
&$+2$ &$-2$ &$-1242$\cred{(-$\frac{69}{7}$)} & $ F'_5$ \\
 &$\left(\frac{-1}{6}\,\frac{-1}{6}\,\frac{-1}{6}\,\frac{-1}{6}\,\frac{-1}{6}\,;
 \frac{-1}{6}\,\frac{-1}{2}\,\frac{1}{6}\,\right)
\left(\frac{-1}{6}\,\frac{-1}{6}\,\frac{-1}{6}\,\frac{-1}{6}\,\frac{-1}{2}\,;
\frac{1}{6}\,\frac{1}{2}\,\frac{1}{2}\,\right)'$ & && &&&&&& &   \\[0.2em]
\hline

$T_{7}^-$
&$\left(\frac{-1}{6}\,\frac{-1}{6}\,\frac{-1}{6}\,\frac{-1}{6}\,\frac{-1}{6}\,;
\frac{1}{3}\,0 \,\frac{-1}{3}\,\right)
\left(\underline{\frac{5}{6}\,\frac{-1}{6}\,\frac{-1}{6}\,\frac{-1}{6}\,}\,\frac{-1}{2}\,;
\frac{-1}{12}\,\frac{1}{4}\,0\,\right)'$ &  $\five'_{5/3}$ & $1$ &+4 &0 &--4 &$-4$
&$-1$ &$-11$  &$+666$\cred{(+$\frac{37}{7}$)} & $F'_6$  \\[0.2em]
&$\left(\frac{-1}{6}\,\frac{-1}{6}\,\frac{-1}{6}\,\frac{-1}{6}\,\frac{-1}{6}\,;
\frac{1}{3}\,0 \,\frac{-1}{3}\,\right)
\left(\frac{1}{3}\,\frac{1}{3}\,\frac{1}{3}\,\frac{1}{3}\,0\,;\frac{-7}{12}\,
\frac{-1}{4}\,\frac{-1}{2}\,\right)'$ & &&&&&&&& &  \\[0.3em]
\hline 
\end{tabular}
\end{center}
\caption{The SU(5)$'$ representations. The red entries are $\Qanom/126$. }\label{tb:SUhfields} }
\end{table}

By the VEV of  $\Phi^{A\alpha}\equiv $\,({\bf 5}$'$,{\bf 2}$'$),
   \begin{eqnarray}
\langle \Phi^{A=3, \alpha=2}\rangle=V_1 \label{eq:Vmix}
\end{eqnarray}
we obtain a group containing two  SU(2)$\times$U(1)$'$ subgroups from non-Abelian factors SU(5)$'\times$SU(2)$'$, \ie a rank-4 subgroup from the rank-5 non-Abelian group, which is denoted as
\begin{eqnarray}
{\rm SU(2)}_1\times{\rm U(1)}_1\times {\rm SU(2)}_2\times{\rm U(1)}_2,
\end{eqnarray}
where the index $i$ is for ${\rm SU(2)}_1$ and the index  $I$ is for ${\rm SU(2)}_2$. In fact, the VEV (\ref{eq:Vmix}) breaks the rank-5 SU(5)$'\times$SU(2)$'$ down to rank-4 SU(4)$'\times$U(1)$'$. The rank-3 SU(4)$'$ is further broken down to rank-2 SU(2)$_1\times$ SU(2)$_2$ by the VEV in the direction,
 \begin{eqnarray}
\langle \overline{\bf 10}_0'\rangle=V_2:   \begin{pmatrix} 0,&0,&0,&0,&0\\
0,&0,&0,&0,&0\\
0,&0,&0,&0,&0\\
0,&0,&0,&0,&V_2\\
0,&0,&0,&-V_2,&0
\end{pmatrix} .  \label{eq:V2}
\end{eqnarray}
Summarizing the above discussion, the rank-5 SU(5)$'\times$SU(2)$'$ is broken down to a rank-3 group  by $V_1$ and $V_2$,
\begin{equation}
{\rm SU(2)_1\times SU(2)_2\times U(1)_Q},\label{eq;GaGroup}
\end{equation}
where 
 \begin{eqnarray}
Q= \begin{pmatrix} \frac{-1}2,&0,&0,&0,&0\\
0,&\frac{-1}2,&0,&0,&0\\
0,&0,&1,&0,&0\\
0,&0,&0,&0,&0\\
0,&0,&0,&0,&0
\end{pmatrix}\otimes  \begin{pmatrix}   +1,&0\\ 0,&-1\end{pmatrix}= Y_1 \otimes  \begin{pmatrix}   +1,&0\\ 0,& -1\end{pmatrix} \label{eq;Q}
\end{eqnarray}
and
  \begin{eqnarray}
 Y_1= \begin{pmatrix} \frac{-1}2,&0,&0,&0,&0\\
0,& \frac{-1}2,&0,&0,&0\\0,&0,&1,&0,&0\\
0,&0,&0,&0,&0\\
0,&0,&0,&0,&0
\end{pmatrix},~  
Y_2= \begin{pmatrix} \frac{+1}{3},&0,&0,&0,&0\\
0,&\frac{+1}{3},&0,&0,&0\\0,&0,&\frac{+1}{3},&0,&0\\
0,&0,&0,&\frac{-1}2,&0\\
0,&0,&0,&0,&\frac{-1}2
\end{pmatrix}.
\end{eqnarray}
Thus, $V_2$ breaks $Y_2$ which does not participate in $Q$ of Eq. (\ref{eq;Q}).
SU(2)$_1$ and SU(2)$_2$ generators are
   \begin{eqnarray}
T^i= \begin{pmatrix} (2\times 2)^i,&0,&0\\
0,&0,&0\\
0,&0,& {\bf 0}
\end{pmatrix},~ T^I= \begin{pmatrix} {\bf 0},&0,&0\\
0,&0,&0\\
0,&0,&(2\times 2)^I
\end{pmatrix}. 
\end{eqnarray}

\begin{table}[!t]
{\tiny
\begin{center}
\begin{tabular}{|c|c|c|c||ccccccc|c|c|}
\hline Sect. & States & SU(2)$'$  & Multiplicity & $Q_1$& $Q_2$ & $Q_3$ & $Q_4$ & $Q_5$ & $Q_6$ &  $Q_{\rm anom}$ & Label \\[0.3em]
\hline\hline

$T_1^0$
&$\left(0\,0\,0\,0\,0\,;\frac{-1}{6}\,\frac{-1}{6}\,\frac{-1}{6}\,\right)
(\underline{1\,0\,0\,0\,}\,0\,; \frac{1}{4}\,\frac{1}{4}\,\frac{1}{2} )'$ &  $(\five',\two')_0$   & $1$ &--2 &--2 &--2 &0 &$+3$ &$-3$ &$-540$\cred{(-$\frac{30}{7}$)} & $F_1'$  \\[0.2em]
&$\left(0\,0\,0\,0\,0\,;\frac{-1}{6}\,\frac{-1}{6}\,\frac{-1}{6}\,\right)
({0\,0\,0\,0\,}\,0\,; \frac{-3}{4}\,\frac{-3}{4}\,\frac{-1}{2} )'$ &&&&&&&&& &   \\[0.5em]
\hline

$T_{1}^0$ &
$\left(0\,0\,0\,0\,0\,;\frac{-1}{6}\,
\frac{-1}{6}\,\frac{-1}{6}\,\right)
\left(0\,0\,0\,0\,1\,;\frac{1}{4}\,\frac{1}{4}\,\frac{1}{2}\,
\right)'$ &  $\two'_0$  & $1$ &$-2$ &$-2$ &$-2$ &0 &$+3$ &$+21$ &$-756$\cred{(-$6$)} & $D_2$  \\[0.5em]
\hline

$T_{1}^+$ &
$\left(\frac{1}{6}\,\frac{1}{6}\,\frac{1}{6}\,\frac{1}{6}\,\frac{1}{6}\,;\frac{1}{3}\,
\frac{-1}{3}\,0\,\right)
\left(\frac{1}{6}\,\frac{1}{6}\,\frac{1}{6}\,\frac{1}{6}\,\frac{1}{2}\,;\frac{1}{12}\,
\frac{3}{4}\,0\,
\right)'$
&$\two'_{-5/3}$  & $1$ &$+4$ &$-4$ &$0$ &$-8$ &$-5$ &$+5$ &$+18$\cred{(+$\frac{1}{7}$)} & $D_3$  \\[0.5em]
\hline

$T_{1}^-$ &
$\left(\frac{-1}{6}\,\frac{-1}{6}\,\frac{-1}{6}\,\frac{-1}{6}\,\frac{-1}{6}\,;\frac{-2}{3}\,
0\,\frac{-1}{3}\,\right)
\left(\frac{1}{3}\,\frac{1}{3}\,\frac{1}{3}\,\frac{1}{3}\,0\,;\frac{-1}{12}\,
\frac{1}{4}\,\frac{1}{2}\,\right)'$
&$\two'_{5/3}$  & $1$ &$-8$ &$0$ &$-4$ &$-4$ &$+5$ &$-5$ &$-774$\cred{(-$\frac{43}{7}$)}  & $D_4$ \\[0.5em]
\hline

$T_{1}^-$ &
$\left(\frac{-1}{6}\,\frac{-1}{6}\,\frac{-1}{6}\,\frac{-1}{6}\,\frac{-1}{6}\,;\frac{1}{3}\,
0\,\frac{2}{3}\,\right)
\left(\frac{1}{3}\,\frac{1}{3}\,\frac{1}{3}\,\frac{1}{3}\,0\,;\frac{-1}{12}\,
\frac{1}{4}\,\frac{1}{2}\,\right)'$
&$\two'_{5/3}$  & $1$ &$+4$ &$0$ &$+8$ &$-4$ &$+5$ &$-5$ &$-270$\cred{(-$\frac{15}{7}$)}  & $D_5$  \\[0.5em]
\hline

$T_{2}^+$ &
$\left(\frac{-1}{6}\,\frac{-1}{6}\,\frac{-1}{6}\,\frac{-1}{6}\,\frac{-1}{6}\,;\frac{1}{6}\,
\frac{-1}{6}\,\frac{1}{2}\,\right)
\left(\frac{1}{3}\,\frac{1}{3}\,\frac{1}{3}\,\frac{1}{3}\,0\,;\frac{1}{6}\,\frac{1}{2}\,0\,
\right)'$ &  $\two'_{5/3}$ & $1$ &$+2$ &$-2$  &$+6$&$-4$ &$-4$ &$-8$  &$-54$\cred{(-$\frac{3}{7}$)} & $D_6$   \\[0.5em]
\hline

$T_{2}^-$ &
$\left(\frac{1}{6}\,\frac{1}{6}\,\frac{1}{6}\,\frac{1}{6}\,\frac{1}{6}\,;\frac{1}{6}\,
\frac{-1}{2}\,\frac{-1}{6}\,\right)
\left(\frac{1}{6}\,\frac{1}{6}\,\frac{1}{6}\,\frac{1}{6}\,\frac{1}{2}\,;\frac{1}{3}\,0\,
\frac{1}{2}\,
\right)'$ &  $\two'_{-5/3}$  & $1$ &$+2$ &$-6$ &$-2$ &$+4$ &$+4$ &$+8$ &$-954$\cred{(-$\frac{53}{7}$)} & $D_7$   \\[0.5em]
\hline

$T_{4}^+$ &
$\left(\frac{1}{6}\,\frac{1}{6}\,\frac{1}{6}\,\frac{1}{6}\,\frac{1}{6}\,;\frac{-1}{6}\,
\frac{1}{6}\,\frac{1}{2}\,\right)
\left(\frac{1}{6}\,\frac{1}{6}\,\frac{1}{6}\,\frac{1}{6}\,\frac{1}{2}\,;\frac{-1}{6}\,
\frac{1}{2}\,\frac{1}{2}\,
\right)'$ &  $\two'_{-5/3}$ & $2$ &--2 &+2 &+6 &$-8$ &$+4$ &$+8$ &$-450$\cred{(-$\frac{25}{7}$)}  & $2D_8$  \\[0.5em]
\hline

$T_{4}^-$ &
$\left(\frac{-1}{6}\,\frac{-1}{6}\,\frac{-1}{6}\,\frac{-1}{6}\,\frac{-1}{6}\,;\frac{-1}{6}\,
\frac{-1}{2}\,\frac{1}{6}\,\right)
\left(\frac{1}{3}\,\frac{1}{3}\,\frac{1}{3}\,\frac{1}{3}\,0\,;\frac{2}{3}\,0\,0\,
\right)'$ &  $\two'_{5/3}$ & $2$ &--2 &--6 &+2 &$+8$ &$-4$ &$-8$ &$-810$\cred{(-$\frac{45}{7}$)} & $2D_9$  \\[0.5em]
\hline

$T_{7}^+$ &
$\left(\frac{1}{6}\,\frac{1}{6}\,\frac{1}{6}\,\frac{1}{6}\,\frac{1}{6}\,;\frac{1}{3}\,
\frac{2}{3}\,0\,\right)
\left(\frac{1}{6}\,\frac{1}{6}\,\frac{1}{6}\,\frac{1}{6}\,\frac{1}{2}\,;\frac{7}{12}\,
\frac{1}{4}\,0\,
\right)'$ &  $\two'_{5/3}$  & $1$ &$+4$ &$+8$ &$0$ &$+4$ &$-5$ &$+5$ &$+1782$\cred{(+$\frac{99}{7}$)} & $D_{10}$ \\[0.5em]
\hline

$T_{7}^+$ &
$\left(\frac{1}{6}\,\frac{1}{6}\,\frac{1}{6}\,\frac{1}{6}\,\frac{1}{6}\,;\frac{-2}{3}\,
\frac{-1}{3}\,0\,\right)
\left(\frac{1}{6}\,\frac{1}{6}\,\frac{1}{6}\,\frac{1}{6}\,\frac{1}{2}\,;\frac{7}{12}\,
\frac{1}{4}\,0\,
\right)'$ &  $\two'_{5/3}$  & $1$ &$-8$ &$-4$ &$0$ &$+4$ &$-5 $ &$+5$ &$-990$\cred{(-$\frac{55}{7}$)} & $D_{11}$  \\[0.5em]
\hline

$T_{7}^-$ &
$\left(\frac{-1}{6}\,\frac{-1}{6}\,\frac{-1}{6}\,\frac{-1}{6}\,\frac{-1}{6}\,;\frac{1}{3}\,0\,
\frac{-1}{3}\,\right)
\left(\frac{-1}{6}\,\frac{-1}{6}\,\frac{-1}{6}\,\frac{-1}{6}\,\frac{-1}{2}\,;\frac{-1}{12}\,\frac{-3}{4}\,0
\right)'$ &  $\two'_{-5/3}$  & $1$ &$+4$&$0$ &$-4$ &$+8$ &$+5 $ &$-5$ &$+234$\cred{(+$\frac{13}{7}$)} & $D_{12}$  \\[0.5em]
\hline 
\end{tabular}
\end{center}
\caption{The SU(2)$'$ representations with the convention of Table \ref{tb:SUhfields}.  We listed only the upper component of SU(2)$'$ from which the lower component can be obtained by applying $T^-$ of SU(2)$'$.  }\label{tb:SU2fields}
}
\end{table}

The SU(2)$_1\times$SU(2)$_2\times
$U(1)$_Q$ quantum numbers are
\begin{eqnarray}
&&\hskip-0.8cm   \Psi_{ij}\oplus\Psi^{i3}\oplus \Psi_{iJ}\oplus \Psi^{I3}\oplus \Psi_{IJ} =({ \bf 1,1})_{+1}   \oplus({\color{green}\bf 2,1})_{\frac{-1}2}   \oplus({ \bf  2,2})_{\frac{+1}2 }\oplus( {\bf 1,2})_{-1}\oplus ({\color{red}\bf 1,1})_0^{(a)}   ,\nonumber\\
&&\hskip-0.8cm  \Phi^{i\alpha}\oplus \Phi^{3\alpha}\oplus \Phi^{I\alpha}= 
 ({\color{green}\bf 2,1})_{\frac{+1}2}+({\bf 2,1})_{\frac{-3}2}+({ \bf 1,1})_{+2}+ ({\color{red} \bf  1,1})_{0}^{(b)} + ({ \bf  \color{blue}1,2})_{+1}+ ({ \color{blue}\bf  1,2})_{-1}   ,\nonumber\\
&&\hskip-0.8cm \psi_{A}=\psi_{i}\oplus \psi_{3}\oplus \psi_{I}= 
({\bf 2,1})_{\frac{+1}2} \oplus({ \bf  1,1})_{-1} \oplus({  \bf 1,2})_{0} , \\
&&\hskip-0.8cm \phi^{\alpha}=  ({\color{green}\bf 1,1})_{+1}+ ({\color{green}\bf  1,1})_{-1},  \nonumber
\end{eqnarray}
where several colored pairs  form vectolike representations.    Removing the green and blue vector-like pairs, and one combination of the red pair $ {\one}_{0,A}= (1/{\sqrt2})[({ \bf  1,1})_{0}^{(a)} -({ \bf  1,1})_{0}^{(b)}]$ where $S(A)$ represents the (anti-)symmetric combination,   we obtain
   \begin{eqnarray}
({ \bf  1,1})_{+1} \oplus ({ \bf 2,2})_{\frac{+1}2 }\oplus ({  \bf  1,2})_{-1}  \oplus ({\color{red} \bf  1,1})_{0,S} \oplus ({\bf 2,1})_{\frac{-3}2} \oplus({ \bf  1,1})_{+2} \oplus
({\bf 2,1})_{\frac{+1}2} \oplus({ \bf  1,1})_{-1} \oplus({  \bf 1,2})_{0}.\label{eq:reps1} 
\end{eqnarray} 
Now, let us break\footnote{In Appendix, we do not break SU(2)$_2$.} SU(2)$_2$ by the VEV $\langle ({\bf 1,2})_{0}\rangle$ which does not carry the $Q$ charge. So, the surviving gauge group is \tdz~where SU(2)$_{\rm dark}$ is  SU(2)$_1$.  Then, there result \tdz~representations  
  \begin{eqnarray}
{\one}_{+1} \oplus 2\cdot  {\bf 2}_{\frac{+1}2 } \oplus 2\cdot {\one}_{-1} \oplus {  \bf  1}_{0} \oplus{\ \bf 2}_{\frac{-3}2} \oplus {\one}_{+2} \oplus { \bf 2}_{\frac{+1}2} \oplus {\one}_{-1}  \oplus 2\cdot {\one}_{0} ,\label{eq:TDZstr} 
\end{eqnarray} 
which are exactly those appearing in Eq. (\ref{eq:MinModel}). 

Considering only the low energy SUSY, we have shown that the minimal chiral model is derivable from a string compactification.
 So, it will be useful if the SUSY scenario is consistent with the unification of gauge coupling constants. Since there are so many unknown parameters in this study,  we deferred  a brief discussion on SUSY scenario to Appendix.

 \section{Conclusion}\label{sec:Conclusion}
We obtained a new chiral model with the gauge group \tdz~without  any gauge and gravitational anomalies.   This gauge group may belong to the dark sector. We also derived this chiral spectrum from a compactification of the heterotic string.  This new chiral theory has a chance to be found at TeV scale accelerators through the kinetic mixing effects. Necessarily, there appear two degrees at low energy: two \darkZ's, or a \darkZ~plus a dark-photon (if $V_S =0$ in Eq. (\ref{eq:gaugeA})).  

  
\acknowledgments{
I thank useful discussions with W.S. Cho and J-C. Park.
This work is supported in part by the National Research Foundation (NRF) grant funded by the Korean Government (MEST) (NRF-2015R1D1A1A01058449) and  the IBS (IBS-R017-D1-2016-a00).}
 
 
\section*{Appendix: Hidden sector SU(5)$'$}
\subsubsection{Mass scales}\label{appA:Masses}

\begin{figure}[!t]
\begin{center}
\includegraphics[width=0.7\linewidth]
{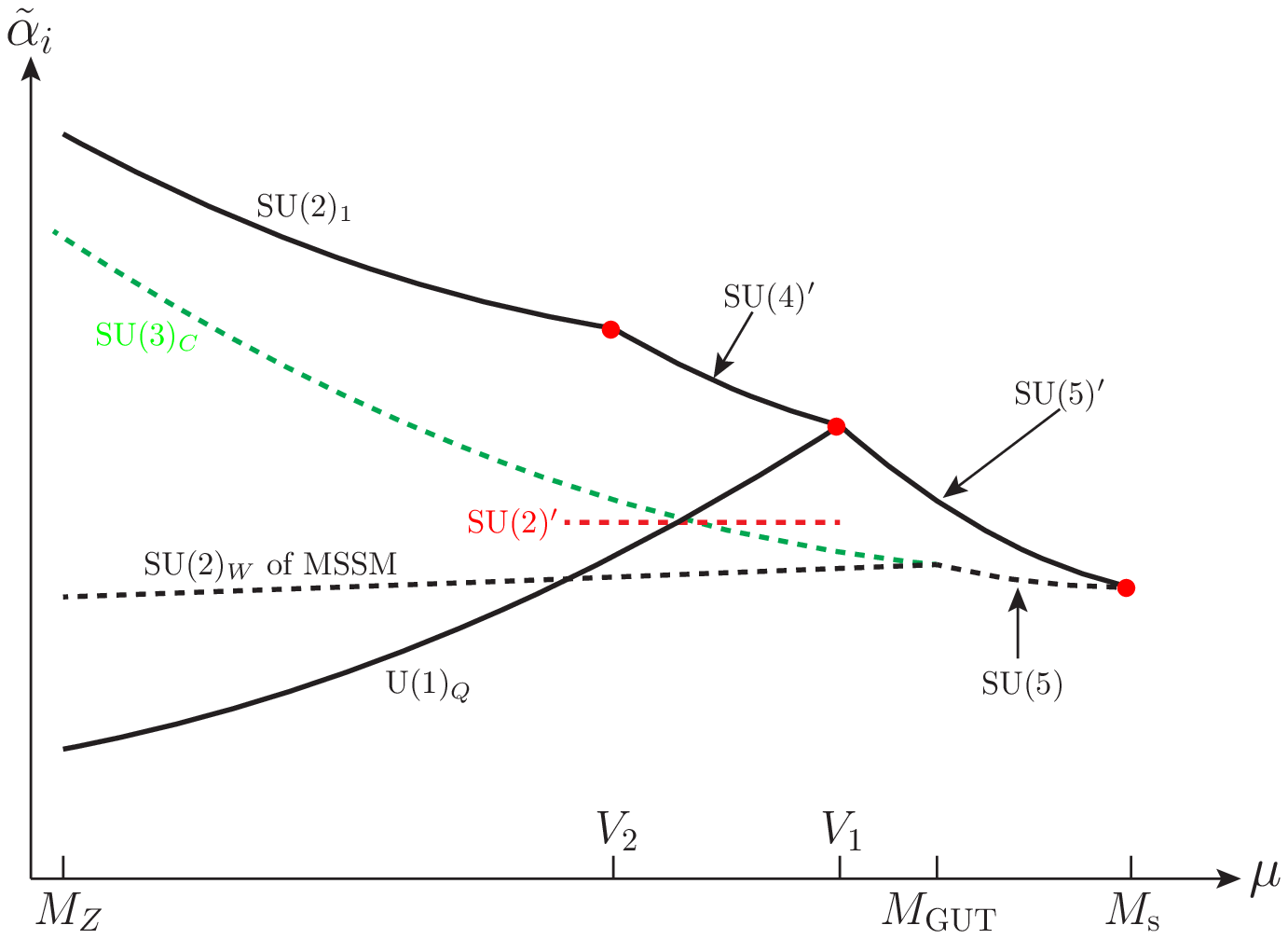}
\end{center}
\caption{The red dashline is for SU(2)$'$.  } \label{fig:Running}
\end{figure}

Toward a suggestion for an ultra-violet completed theory, we discuss at which scales symmetry breakings are introduced. Firstly, we need one confining force for dynamical SUSY breaking \cite{Nilles83,DIN85}. The confining non-Abelian gauge group at the intermediate scale is chosen as SU(5)$'$. Around the same scale,  SU(5)$'$ is broken down to SU(4)$'$ and at a somewhat lower scale to SU(2)$_1$ by the condensation of matter superfield, breaking SU(2)$_2$.   Because SU(2)$_2$ is neutral under the \tdz~transformation, the discussion leading to the minimal model is intact. A rough sketch of related scales is shown   in Fig. \ref{fig:Running}. 

The confining superfields in Eq. (\ref{eq:reps1}) are
 \begin{eqnarray}
 ({ \bf 2,}A^\alpha)_{\frac{+1}2 }\oplus ({\bf  1},B^\alpha)_{-1}    \oplus({\bf 1},C^\alpha)_{0},\label{eq:Strong} 
\end{eqnarray} 
where $\alpha=\{1,2\}$ counts the number of color degrees of SU(2)$_2$. The  anomaly matching conditions \cite{Hooft79} must lead to the following composite states under \tdz,
\dis{
2\cdot  {\bf D}_{\frac{+1}2 } \oplus 2\cdot {\bf S}_{-1},
}
where the composite states {\bf D} and {\bf S} are \tdz~doublets and singlets, respectively, composed of $A,B$, and $C$ degrees,
\dis{
{\bf D}_{\frac{+1}2 } \propto \epsilon_{\alpha\beta}({ \bf 2},A^\alpha)_{\frac{+1}2 }({\bf 1},C^\beta)_{0},\qquad {\bf S}_{-1} \propto 
\epsilon_{\alpha\beta}({\bf 1},B^\alpha)_{\frac{+1}2 }({\bf 1},C^\beta)_{0}.
}
Even though SU(2)$_2$ is smaller than the color SU(3)$_C$, it can confine at the intermediate scale if SU(5)$'$ and SU(4)$'$ runs between the GUT scale and the intermediate scale. So, the SU(4)$'$ breaking VEV $V_2$, Eq. (\ref{eq:V2}), is around the intermediate scale.

\dis{
{\rm SU(5)}' \times{\rm SU(2)}' \big|_{V_1< \Mg} \longrightarrow
{\rm SU(4)}' \times{\rm U(1)}' \big|_{V_2} \longrightarrow
{\rm SU(2)}_1  \times{\rm SU(2)}_2 \times{\rm U(1)}_Q \big|_{\Mi} \longrightarrow {\rm SU(2)_{\rm dark}\times U(1)_Q}
}
where $V_1< \Mg$. From the compactification scale down to \Mgt, SU(5)$'$ runs more steeply than SU(2)$'$,   which is illustrated as the separate couplings at $V_1$  in Fig. \ref{fig:Running}.

In the radiative breaking of the SM gauge group in the MSSM, the large top quark Yukawa coupling plays a crucial role. To break \tdz, near the electroweak scale, we need some large Yukawa coupling(s) involving $ { \bf 2}_{\frac{+1}2}$'s, ${ \bf 2}_{\frac{-3}2}, {\one}_{-1}$'s, and ${\one}_{+2} $ in Eq. (\ref{eq:TDZstr}).

\subsubsection{Running of couplings}
\label{app:Running}

For a rough guess of the coupling constants, we use just one loop evolution equations.
With the mass order of  Fig. \ref{fig:Running}, we have the following running of gauge couplings,\footnote{Spectra of \SUflip~are counted from Ref. \cite{KimKyae17}.  The gauge group U(1)$_Q$, which is not the anomalous U(1), survives down to the TeV scale.}
\dis{
SU(5)_{\rm flip}:\frac{1}{g_5^2(\Mg)}=\frac{1}{g_5^2(M_{\rm st})}+\frac{1}{8\pi^2}\left(-3\cdot 5+ 12\right)\ln\frac{M_{\rm st}}{\Mg}, \label{eq:flipRun}
}
\dis{
SU(5)' :\frac{1}{ \tilde{g}_4^2(V_1)}=\frac{1}{ \tilde{g}_5^2(M_{\rm st})}+\frac{1}{8\pi^2}\left(-3\cdot 5+ 3\right)\ln\frac{M_{\rm st}}{V_1}, \label{eq:hidRun}
}
\dis{
SU(4)' :\frac{1}{ \tilde{g}_4^2(V_2)}=\frac{1}{ \tilde{g}_4^2(V_1)}+\frac{1}{8\pi^2}\left(-3\cdot 4+ 4\right)\ln\frac{V_1}{V_2}, 
}

\dis{ SU(2)_1:\frac{1}{ \tilde{g}_2^2(M_Z)} =\frac{1}{ \tilde{g}_2^2(V_1)}+\frac{1}{8\pi^2}\left(-3\cdot 2+ 4\right)\ln\frac{V_1}{M_Z},}
\dis{
U(1)_Q:\frac{1}{ \tilde{g}_Q^2(M_Z)}=\frac{1}{ \tilde{g}_Q^2(V_1)}+\frac{1}{8\pi^2}\left(+ 4\right)\ln\frac{V_1}{M_Z},
}
where SU(5)$'$ couplings are tilded.
In the figure, we also sketched the running of the SU(2)$_W$ ( SU(3)$_C$) coupling as the (green) dash line. From the observed value of $\alpha_2$ at $\mu=M_Z$ for $\sin^2\theta_W(M_Z)\simeq 0.23$ \cite{KimRMP81}, we obtain its GUT scale value. Identifying this as the SU(5)$'$ coupling at \Mgt, we estimate the couplings as sketched in   Fig. \ref{fig:Running}. We assume that SU(4)$'$ gauginos condenses at $\approx V_1$, 
\dis{
\langle  \tilde{G}_{A}^B\tilde{G}^{A}_B\rangle\ne 0.
}
Then, between $V_1$ and $V_2$, we consider the group SU(4)$'$. At $V_2$, $\tilde{\alpha}_2 $ has not reached to an order one value, but   $\langle \Psi_{IJ}\rangle$ can be developed at $V_2$. The representation {\bf 6}  of SU(4)$'$ has a larger Casimir operator $\frac{5}{2}$ than that of the fundamental representation $\frac{15}{8}$.  So, we expect that there appear composites ${\bf D}_{\frac{+1}{2}}$ and  ${\bf S}_{-1}$ discussed in Appendix A.  

Using the electroweak coupling at $M_Z$, $\alpha_2\simeq 3.38\times 10^{-2}$ \cite{Earler14}, we obtain its evolution to \Mgt, $\alpha_2(\Mg)\simeq 0.0412$ where $\Mg=2.5\times 10^{16\,}\gev$ is used. At the hypothetical string scale $M_s\simeq 0.7\times 10^{18\,}\gev$, we obtain the coupling $\alpha_5(M_s)\simeq 0.0389$ which is equated to $\tilde{\alpha}_5 (M_s)$. Now, we can run the hidden sector couplings down from $M_s$. Suppose that gaugino condensation occurs at $M_{\rm cond}=10^{13\,}\gev$. With $V_1=\Mg$,
\dis{
 {\rm SU(5)'}:~\tilde{\alpha}_5(\Mg)\simeq 0.0517,~{\rm for}~\Mg=2.5\times 10^{16\,}\gev.\label{al5Coupl}
}
For SU(5), the Casimir of the adjoint representation   is 25/12 times larger than that of the fundamental representation. So, gauginos couple more strongly than the fundamentals.   The SU(4)$'$ coupling at $V_2$ is
\dis{
 &{\rm SU(4)'}:~\tilde{\alpha}_4(V_2)\simeq 0.1066,~{\rm if}~V_2=10^{13\,}\gev.\label{al4Coupl}
}
Let us equate (\ref{al4Coupl}) as   the SU(2)$_1$ coupling at $V_2$. Then, the SU(2)$_1$ and U(1)$_Q$ couplings at $M_Z$ are
\dis{
   &{\rm SU(2)_1}:~\tilde{\alpha}_2(M_Z)\simeq 0.743,\\
   &{\rm U(1)}_Q:~\tilde{\alpha}_Q(M_Z)\simeq 0.0251, \label{eq:SU1Couplings}
}
such that the mixing angle is $|\sin\theta|_{M_Z}\simeq \sqrt{\tilde{\alpha}_Q / (\tilde{\alpha}_2 
+\tilde{\alpha}_Q})|_{M_Z}\simeq  0.213$.
At $M_Z$, $\tilde{\alpha}_2(M_Z)$ is  much larger than the electroweak coupling $\alpha_2(M_Z)$. If the VEV $\langle {\two}_{{1}/{2}}\rangle$ has the same order as $V_D$ of Eq. (\ref{eq:gaugeZ}), then we obtain \darkZ$_1$ mass at the electroweak scale. Actually, the \darkZ~masses depend on the parameters, the mixing angle and the VEVs given in Eqs. (\ref{eq:gaugeZ}) and (\ref{eq:gaugeA}).
  
We also  note that there exists a possible superpotential term,
 \begin{eqnarray}
  \Psi_{IJ}\Phi^{I\alpha}
\Phi^{J\beta}\epsilon_{\alpha\beta}\sim
{\color{red}({\one,\one})_0 }\,{\color{blue}({\one,\two})_{+1}({\one,\two})_{-1} }
  \end{eqnarray}
which may allow  $\langle\Psi_{IJ}\rangle=V_2$ by the condensation of $\langle{({\one,\two})_{+1}({\one,\two})_{-1} }\rangle$.

\subsubsection{Confinement of SU(2)$_1$}
\label{appA:Confinement}
The large SU(2)$_1$ coupling in Eq. (\ref {eq:SU1Couplings}) suggests a possibility that  SU(2)$_1$ confines around the electroweak scale. Let us consider four doublets of Eq. (\ref{eq:MinModel}) as
\begin{eqnarray}
D_1= \begin{pmatrix}P_1\\ N_1\end{pmatrix}_{L,\frac{+1}2} ,~D_2= \begin{pmatrix} P_2\\ N_2 \end{pmatrix}_{L,\frac{+1}2} ,~\bar{D}_1= \begin{pmatrix} P_2\\ N_2 \end{pmatrix}_{R,\frac{-1}2} ,~\bar{{\cal D}}_2= \begin{pmatrix} P_2\\ N_2 \end{pmatrix}_{R,\frac{+3}2} 
\end{eqnarray}
where two SU(2)$_1$ doublets are represented as R-handed chiral fields and the subscripts are the U(1)$_Q$ charges. Below the SU(2)$_1$ confinement scale,  we consider the following condensations 
\dis{
\langle \bar{D}_1 D_1\rangle=V_D,~\langle \bar{\cal D}_2 D_2\rangle=V_S,
}
and use the mass ratio presented in Eq. (\ref{eq:ratio}). When $\tilde{\alpha}_2$ becomes order 1 at the scale $\mu_2$, let us assume that SU(2)$_1$ confines. The condenstation scale is guessed as $\mu_2/2$, following the estimate of the QCD condensation scale $\langle \bar{u}u\rangle \approx1\,\gev/3$ where $\alpha_C(1\,\gev)\simeq O(1)$. $\tilde{\alpha}_2$ becomes order 1 at $\mu_2\simeq 22.2\,\gev$. In this setup, we estimate the masses of two \darkZ's as
\cite{Susskind79},
\dis{
&M_1\sim \frac{\tilde{g}_2(\mu_2)}{2}\,\frac{\mu_2}{2}\gtrsim 20\,\gev,\\[0.3em]
&M_2\simeq \sqrt2\sin\theta\frac{V_S}{V_D}\sim 6\,\gev,
}
where we used $V_S=V_D$ and $\sin\theta=0.2$.
Note, however, that our estimate is very primitive because we used $V_1=\Mg$, one-loop running for gauge coupling evolution, followed the hypothetical SUSY breaking, and a naive chiral symmetry breaking below the SU(2)$_1$ confinement scale. Nevertheless, this crude estimate has lead to two electroweak scale \darkZ's.


\end{document}